\documentclass[a4paper,11pt]{article}
\usepackage{pos}

\title{Exploring the Most Extreme Blazars: New Insights from MAGIC}
\ShortTitle{Extreme Blazars with MAGIC}

\author*[a]{L. Foffano} 
\author[b]{C. Arcaro}
\author[c]{A. Arbet-Engels}
\author[a]{F. D’Ammando}
\author[a]{P. Da Vela}
\author[e]{J. Jormanainen}
\author[f]{D. Linder}
\author[a]{S. Menon}
\author[b]{E. Prandini}
\author[d]{S. Ventura}
\author[g]{E. Visentin,}
\onbehalf{for the MAGIC Collaboration\\ } 

\affiliation[a]{National Institute for Astrophysics (INAF), I-00136 Rome, Italy}
\affiliation[b]{Universit\`a di Padova and INFN, I-35131 Padova, Italy}
\affiliation[c]{Max-Planck-Institut f\"ur Physik, D-85748 Garching, Germany}
\affiliation[d]{Universit\`a di Siena and INFN Pisa, I-53100 Siena, Italy}
\affiliation[e]{Finnish MAGIC Group: Finnish Centre for Astronomy with ESO, Department of Physics and Astronomy, University of Turku, FI-20014 Turku, Finland}
\affiliation[f]{ETH Z\"urich, CH-8093 Z\"urich, Switzerland}
\affiliation[g]{INFN Sezione di Torino and Universit\`a degli Studi di Torino, I-10125 Torino, Italy}

\emailAdd{luca.foffano@inaf.it}

\abstract{
Extremely high-peaked BL Lac objects (or extreme blazars) are unique extragalactic laboratories where particle acceleration processes are pushed at their physical limits. In these blazars, synchrotron emission peaking above keV energies is reprocessed to very-high-energy (VHE, energies > 100 GeV) gamma rays, often resulting in very hard TeV spectra. Over the past two decades, they have attracted a growing interest from the scientific community, both experimentally and theoretically, as crucial targets for understanding the diversity within the blazar class.
On the experimental side, new sources have been detected and characterized, populating the extreme blazars class. Notably, VHE campaigns have revealed evidence of emerging spectral differences in this energy band, suggesting inhomogeneity within this class of sources. Recent studies have also unveiled intriguing differences in the temporal evolution of their spectral emission. On the theoretical side, these spectral differences are challenging the current standard emission and acceleration models for blazars, suggesting the need for more complex theoretical frameworks.
In this contribution, we present the latest results from recent MAGIC Collaboration observing campaigns aimed to enlarge the extreme blazars population at VHE and understand the origin of their extreme properties. Furthermore, we will present the results of the most recent observations, discussing analogies and differences with well-known sources such as the archetypal 1ES 0229+200, as well as interpretations of their non-conventional spectral emission.
}

\ConferenceLogo{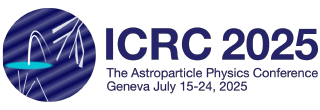}

\FullConference{39th International Cosmic Ray Conference (ICRC2025)\\
 15–24 July 2025\\
Geneva, Switzerland\\}

\begin{document}
\maketitle

\section{Introduction}
\noindent
Among the large class of active galactic nuclei, blazars are characterized by a relativistic jet pointing directly towards to the observer, making them valuable sources to directly explore their particle acceleration mechanisms.

Blazars constitute an heterogeneous population of sources, subdivided in subclasses based on their radiation efficiency \citep[e.g.][]{Fossati98}. 
BL Lac objects are blazars with lower environmental particle density, which implies a lower luminosity and higher particle acceleration efficiency. 
The most energetic sources of this class are called \textit{extreme blazars} \citep[e.g.][]{Costamante:2001pu}, a short name indicating Extremely High-Peaked BL Lac objects (EHBLs).  Their spectral energy distribution (SED) typically displays two emission components. Extreme blazars are defined on the basis of their synchrotron peak energy, which lies above $10^{17}$~Hz ($\sim0.3$ keV) in the X-ray band \citep{2010ApJ...716...30A}. 
This low-energy emission is then reprocessed and produces a bump of radiation in the SED extending deeply into very high-energy (VHE, E$>$100 GeV) gamma-ray band, whose origin is attributed mainly to inverse Compton emission of the synchrotron photons \citep[SSC emission, ][]{Maraschi:SSC, Tavecchio:SSC} with possible additional contributions of further leptonic-hadronic components \citep[e.g.,][]{Costamante2018}.

These extreme energies - which are optimally investigated by the MAGIC telescopes - turn out to be key to understanding the physical processes behind their extreme properties and to test them as cosmological probes \citep[e.g.,][]{2020NatAs...4..124B}. In fact, at TeV energies, the extreme blazar population appears to be divided into heterogeneous classes of sources \citep{foffano2018}. Some of them show extremely hard TeV spectra - named \textit{hard-TeV} extreme blazars - while others show softer spectra. Some specific sources have experienced large flux variations,  showing temporarily characteristics of extreme blazars during flaring episodes, for example in the case of 1ES~2344+512 \citep{2020A&A...640A.132M, 2024A&A...682A.114M} and 1ES~1959+650 \citep{2020A&A...638A..14M}.

The interpretation of such exotic and various spectral properties is still challenging within the existing theoretical models, requiring extreme parameter values.
The growth of their population represent a key to improve our understanding on the physical processes behind their extreme properties.


\section{Extreme blazars at TeV gamma-ray energies}
\noindent
The low radiation efficiency of extreme blazars - combined with their enhanced particle acceleration efficiency and diverse possible spectral properties - makes these objects hard to be detected at TeV energies.
To address the search for new TeV extreme blazars, systematic observations  are carried out on the basis of a list of promising targets. Such candidates are selected among catalogs of blazars based on a complex set of observational criteria - such as hardness of the spectrum in X-rays and gamma~rays, estimated redshift, and multi-wavelength correlations supporting their classification as extreme blazars  - which make these sources likely detectable at TeV energies with the current generation of Cherenkov telescopes. 
As a result, the population of known TeV extreme blazars is steadily growing, driven by the ongoing discovery of new sources in this energy range.

\section{A new catalog of MAGIC extreme blazars}
\noindent
The MAGIC telescopes are a system of two 17-m diameter imaging atmospheric Cherenkov telescopes (IACTs) located at an altitude of 2200 m above sea level on the island of La Palma, in Spain. They sensitivity is optimized for gamma-ray photons with energies ranging from about 20~GeV to 100~TeV   \citep{magicperf_1:2015, magicperf_2:2015}.

In this contribution, we present the results of the long-standing observing program led by the MAGIC Collaboration dedicated to the search for new TeV extreme blazars, aimed at extending their population and characterizing their emission behavior.
This program has already proven successful, providing to the community several new EHBL detections at TeV energies. 

This study allows us to build a new catalog of 11 sources, based on MAGIC observations carried out between 2017 and 2025, investigating their flux variability and spectral properties. Among them, we provide new long-term monitoring data of 3 known extreme blazars, extending the findings of  our previous publication \citep{1st_MAGIC_EHBL_catalog}. In addition, the catalog includes -- for the first time -- data on 8 further sources, providing new TeV detections.
Complementary multi-wavelength observations in the optical-UV and X-ray bands were carried out with dense temporal and spectral coverage. This rich dataset allows us to investigate and test various theoretical scenarios describing the underlying emission processes.
The complete set of results and the comparison among this various set of sources allow us to discuss interesting physical processes of the category of extreme blazars.

\section*{Acknowledgements}
\noindent
We would like to thank the Instituto de Astrof\'{\i}sica de Canarias for the excellent working conditions at the Observatorio del Roque de los Muchachos in La Palma. The financial support of the German BMBF, MPG and HGF; the Italian INFN and INAF; the Swiss National Fund SNF; the grants PID2019-107988GB-C22, PID2022-136828NB-C41, PID2022-137810NB-C22, PID2022-138172NB-C41, PID2022-138172NB-C42, PID2022-138172NB-C43, PID2022-139117NB-C41, PID2022-139117NB-C42, PID2022-139117NB-C43, PID2022-139117NB-C44, CNS2023-144504 funded by the Spanish MCIN/AEI/ 10.13039/501100011033 and "ERDF A way of making Europe; the Indian Department of Atomic Energy; the Japanese ICRR, the University of Tokyo, JSPS, and MEXT; the Bulgarian Ministry of Education and Science, National RI Roadmap Project DO1-400/18.12.2020 and the Academy of Finland grant nr. 320045 is gratefully acknowledged. This work was also been supported by Centros de Excelencia ``Severo Ochoa'' y Unidades ``Mar\'{\i}a de Maeztu'' program of the Spanish MCIN/AEI/ 10.13039/501100011033 (CEX2019-000920-S, CEX2019-000918-M, CEX2021-001131-S) and by the CERCA institution and grants 2021SGR00426 and 2021SGR00773 of the Generalitat de Catalunya; by the Croatian Science Foundation (HrZZ) Project IP-2022-10-4595 and the University of Rijeka Project uniri-prirod-18-48; by the Deutsche Forschungsgemeinschaft (SFB1491) and by the Lamarr-Institute for Machine Learning and Artificial Intelligence; by the Polish Ministry Of Education and Science grant No. 2021/WK/08; and by the Brazilian MCTIC, CNPq and FAPERJ.

\bibliographystyle{unsrtnat}
\bibliography{biblio_2nd_MAGIC_EHBL_catalog}{}

\end{document}